\documentclass{article}

\usepackage{PRIMEarxiv}

\usepackage[utf8]{inputenc} 
\usepackage[T1]{fontenc}    
\usepackage[hyphens,spaces,obeyspaces]{url}
\usepackage{hyperref}
\usepackage{booktabs}       
\usepackage{amsfonts}       
\usepackage{nicefrac}       
\usepackage{microtype}      
\usepackage{lipsum}
\usepackage{fancyhdr}       
\usepackage{graphicx}       
\usepackage{tabularx}
\usepackage{multirow}

\usepackage{tikz}
\usetikzlibrary{shapes.geometric, arrows}
\usepackage{pgfplots}
\pgfmathdeclarefunction{gauss}{2}{%
\pgfmathparse{1/(#2*sqrt(2*pi))*exp(-((x-#1)^2)/(2*#2^2))}%
}
\usepackage{pgfplots}  
\pgfplotsset{width = 0.75\linewidth ,compat=1.17} 

\title{Towards Assessing Data Bias in Clinical Trials
}

\author{
  Chiara Criscuolo \\
  Politecnico di Milano \\
  Milan, Italy\\
  \texttt{chiara.criscuolo@polimi.it}
\And
  Tommaso Dolci \\
  Politecnico di Milano \\
  Milan, Italy\\
  \texttt{tommaso.dolci@polimi.it}
\And
  Mattia Salnitri \\
  Politecnico di Milano \\
  Milan, Italy\\
  \texttt{mattia.salnitri@polimi.it}
}

\begin{document}
\maketitle

\begin{abstract}
Algorithms and technologies are essential tools that pervade all aspects of our daily lives. In the last decades, health care research benefited from new computer-based recruiting methods, the use of federated architectures for data storage, the introduction of innovative analyses of datasets, and so on. Nevertheless, health care datasets can still be affected by data bias. Due to data bias, they provide a distorted view of reality, leading to wrong analysis results and, consequently, decisions. For example, in a clinical trial that studied the risk of cardiovascular diseases, predictions were wrong due to the lack of data on ethnic minorities. It is, therefore, of paramount importance for researchers to acknowledge data bias that may be present in the datasets they use, eventually adopt techniques to mitigate them and control if and how analyses results are impacted.

This paper proposes a method to address bias in datasets that: \emph{(i)} defines the types of data bias that may be present in the dataset, \emph{(ii)} characterizes and quantifies data bias with adequate metrics, \emph{(iii)} provides guidelines 
to identify, measure, and mitigate data bias for different data sources. The method we propose is applicable both for prospective and retrospective clinical trials. We evaluate our proposal both through theoretical considerations and through interviews with researchers in the health care environment.
\end{abstract}


\section{Introduction}\label{sec:intro}
In the last decades evidence-based medicine relied consistently on large datasets.
New technologies, such as data mining and data lakes, allow to analyze larger and larger datasets obtaining astonishing results that led to new and more reliable therapies~\cite{beam2018big}, such as in the case of diabetic retinopathy detection~\cite{gulshan2016development,krause2018grader}. Yet, reliable results directly depend on high-quality datasets. Biased datasets, i.e., dataset that do not correctly represent the population, inevitably lead to distorted results. This is particularly important for evidence-based medicine, where wrong or not effective therapies might be applied as a consequence of wrong analyses. For instance, in the controversial case of IBM Watson Oncology system, machine learning algorithms return erroneous recommendations whenever the training of the system is not well planned~\cite{grote2020ethics}. The identification and measurement of bias in health care dataset is, therefore, essential to have reliable results and high-quality health care. 

Even though bias and its impacts are well studied in specific research fields and application sectors of Computer Science, such as automated-decision making for hiring~\cite{lambrecht2019algorithmic}, justice~\cite{angwin2016machine} or image recognition~\cite{papakyriakopoulos2021beyond}, 
they are not frequently considered for health care datasets. Unluckily, information on bias is not provided with these datasets, leaving researchers and doctors clueless on the possible impacts on their analysis and with no means to mitigate them~\cite{char2018implementing}. In order to get the benefits of event-based medicine and, therefore, to provide the best possible health care, information bias need to be considered as an essential part of datasets. This issue affects both retrospective clinical trials, which use already collected datasets that may be affected by data bias unknown at the time of data collection, yet relevant for the current trial, and prospective clinical trials, where datasets are collected for a specific clinical experiment.

This paper provides an insightful definition of the types of data bias that may affect health care datasets for retrospective clinical trials, and a method that will help health care researchers in: (i) identifying and measuring bias for health care datasets; (ii) allowing them to filter datasets based on bias information that will be stored on metadata; (iii) choosing the appropriate mitigation action to reduce bias before using the dataset. This method can be used for perspective clinical trials by monitoring the data collection and guide researchers towards the mitigation of identified bias. To reach these objectives, this paper defines a set of guidelines that will guide researcher on the application of the method. The method proposed in this paper has been validated through interviews with health care researchers and experts in bioethics and philosophy.

The paper is structured as it follows. Section~\ref{sec:context} gives an overview of the context of bias and health care datasets. Section~\ref{sec:preliminaries} describes the baseline of this paper, while Section~\ref{sec:soa} provides the most relevant research work on data bias in health care datasets. Sections~\ref{sec:method} and~\ref{sec:bias} are the contribution of this paper: they describe the method proposed, the types of data bias involved, measurements and mitigation actions. Section~\ref{sec:validation} illustrates the initial validation of the method proposed.
Section~\ref{sec:conclusion} concludes the paper.

\section{Data Bias in Health Care}\label{sec:context}

Recently, in the context of health care research, the employment of new information technologies is steadily increasing, driven by the admirable goal of improving effectiveness and efficiency of various aspects of medicine. For instance, machine learning proved to be extremely effective when used for detecting diabetic retinopathy~\cite{gulshan2016development,krause2018grader} and skin cancer~\cite{esteva2017dermatologist}, with results equal to or greater than those of human physicians. Combining big data and machine learning algorithms has been shown to significantly improve the accuracy of analyses~\cite{beam2018big}.

However, despite the enormous advantages of using computer technologies in health care, serious ethical issues emerged on several occasions. For example, it has been shown that systems for algorithmic decision-making for medical diagnosis and treatment recommendations, despite improving accuracy, also increase opacity and uncertainty, especially when using deep learning techniques~\cite{grote2020ethics,topol2019high}. Opacity largely affects aspects such as accountability and fairness: famous is the case of IBM Watson Health, whose algorithms for cancer treatment recommendation turned out to be erroneous due to the small amount and the synthetic nature of the training data~\cite{grote2020ethics}.
In this context, many of the emerged risks are related to data used to train the model or validate the underlying algorithms. In particular, the presence of bias in the data heightens the risk of unfairness and disparity.
For instance, whenever a model is used for patients whose data differ from the data on which the model has been trained, results of the clinical trial are biased~\cite{biasHealthCare}. Also, bias may arise when the study population is not well selected~\cite{gerhard2008bias}.

In recent years, many ethical risks concern ethnic groups, which are often under-represented in clinical trials. Among the problems were the systematic misclassification of variants as pathogenic for Africans~\cite{manrai2016genetic} and the inconsistent performance of risk calculators for cardiovascular disease in multi-ethnic studies, especially for ethnic minorities such as African Caribbeans and South Asian women~\cite{tillin2014ethnicity}. Additionally, new frontiers of medicine such as genome sequencing tend to exclude ethnic minorities in a worrying way, as evident from the fact that by 2009 fewer than one percent of genome investigations included Africans~\cite{wapner2018cancer}.

\section{Preliminaries}\label{sec:preliminaries}

Considerable research attention is being paid to aspects of responsible data analysis and \emph{Data Science Ethics}, which lays the foundation for data bias analysis. 
Data Science Ethics highlights the need for ethical analyses, focusing on concepts such as fairness, diversity, accountability, transparency and privacy. It analyzes the meaning and nature of computational operations, the interactions between hardware, software and data, considering the variety of digital technologies that make them possible. Data Science Ethics emphasizes the complexity of the ethical challenges posed by data science. The rest of this section gives definitions of fundamental concepts used in the rest of the paper: \emph{Bias}, \emph{Fairness} and \emph{Diversity}.

\textbf{Bias} is defined as an ``inclination, or prejudice for, or against one person, or group, especially in a way considered to be unfair"~\cite[p.4]{pitoura2020social}. Sometimes bias is \emph{desirable} and part of the correct system functionality. For example, in a system that predicts the likelihood of a person to have a specific disease, the group of individuals who actually has the disease should be systematically attributed a higher probability than the group who does not have it. An \emph{undesired bias} is ``a bias that is considered problematic, possibly unfair by the stakeholders of the system or other persons impacted by the system''~\cite[p.3]{dataManagementPipeline}. Typically, this is observed when bias relates to a protected attribute, which is defined as ``a sensitive attribute for which non-discrimination should be established''~\cite[p.2]{verma2018fairness}. In the last few decades, many studies focused on the types of data bias that can be present in data \cite{mehrabi2021survey,pitoura2020social}.
 
\textbf{Fairness}, in the context of decision-making, is related to the concept of ``absence of any prejudice or favoritism toward an individual or a group based on their inherent or acquired characteristics"~\cite[p.100]{DBLP:journals/ai/SaxenaHDRPL20}. 
It is also associated with the so-called “lack of bias”~\cite{stoyanovich2016data}. However, there is no universal definition of fairness, mainly because it is a broad concept that intersects different scenarios, thus in literature there are many formulations. To assess fairness there exist many formal metrics~\cite{verma2018fairness}, such as statistical ones that evaluate the distribution of positive predictions among instances in the dataset.

Finally, \textbf{diversity} is ``a general term used to capture the quality of a collection of items, or of a composite item, with regards to the variety of its constituent elements"~\cite[p.1]{Drosou2017DiversityIB}. There are many metrics to quantify diversity; some of them are based on the notion of distance and rely on pair-wise similarity between elements, while others are based on coverage, that measures the extent to which the elements cover a predeﬁned number of aspects~\cite{Drosou2017DiversityIB}.

\section{Related Work}\label{sec:soa}

In this section, we separate related work concerning methods for addressing data bias in generic datasets using Computer Science techniques, and, in the second part, methods that address data bias in health care. 

\subsection{Data Bias in Computer Science}

Most of the works in Computer Science, and specifically in Data Science Ethics, focus on ensuring fairness, generally by algorithmic data bias discovery and mitigation in prediction tasks, more specifically for classification algorithms \cite{adebayo2016fairml,bellamy2018ai,tramer2017fairtest}.

There are three possible approaches to enforce fairness in data analysis applications: \emph{(i) pre-processing techniques}, i.e., procedures to verify that the training data are fair before the application of the algorithm; \emph{(ii) in-processing techniques}, i.e., procedures to ensure that, during the learning phase, the algorithm does not inherit the bias present in the data, and \emph{(iii) post-processing techniques}, i.e., procedures to correct the algorithm predictions, and consequently decisions, and make them fairer.

In this context, one of the most notable work is \emph{AI Fairness 360: An Extensible Toolkit for Detecting, Understanding, and Mitigating Unwanted Algorithmic Bias} \cite{bellamy2018ai}, an open-source framework to reach algorithmic fairness for classifiers.
AI Fairness 360 quantifies and mitigates data bias using a variety of statistical measures, exploiting pre-processing, in-processing and post-processing techniques.

A work that studies both fairness and diversity is \emph{Nutritional Labels for Data and Models} by Stoyanovich and Howe \cite{DBLP:journals/debu/StoyanovichH19}.
The authors develop an interpretability and transparency tool based on the concept of \emph{Nutritional Labels}, drawing an analogy to the food industry, where simple and standardized labels convey information about ingredients and production processes. 
\emph{Ranking Facts} \cite{DBLP:journals/debu/StoyanovichH19} is a system based on nutritional labels that visually expresses fairness through statistical measures, and diversity through the distribution for each category.

One more recent work is \cite{asudeh2019assessing}, a system to assess diversity, and more specifically coverage, for a given dataset over multiple categorical attributes. Based on new efficient techniques to identify the regions of the attribute space not adequately covered by data, this system can determine the least amount of data that must be added to solve the lack of diversity.

\subsection{Data Bias in Health Care}

In the area of health care, the majority of work focuses on theoretical and philosophical aspects, and only few of them give more precise guidelines about data bias identification and measurement.

A famous work is \cite{cohen2014legal}: the authors analyze major challenges when implementing predictive analytics in health care settings. They make broad recommendations for overcoming challenges raised in the four phases of the lifecycle of a predictive analytics model: acquiring data to build the model, building and validating it, testing it in real-world settings, and disseminating and using it more broadly. According to the authors, it is essential for predictive analytics models to be constantly evaluated, updated, re-implemented, and reevaluated, because the presence of bias threatens the trust of patients, providers and the public. Nonetheless, the authors remain optimistic that predictive analytics can help build stronger and more dynamic systems.

Another notable work is \cite{biasHealthCare}, that presents a conceptual framework of how different types of bias relates to one another. From our point of view, it is particularly relevant the classification of bias that arise from machine learning algorithms in the health care area. The authors provide a list of data bias definitions divided into four categories: \emph{bias in model design}, \emph{bias in training data}, \emph{bias in interactions with clinicians} and \emph{bias in interactions with patients}.
They also reveal that the interactions of model predictions with clinicians and patients may exacerbate health care disparities.

As far as we know, there is no method in literature that specifically guides the user in searching and measuring data bias in the health care context, using the instruments from Computer Science and particularly Data Science Ethics.

\section{Method}\label{sec:method}

We propose a framework that consists in a pipeline for identifying and measuring bias in health care datasets. This work targets health care researchers and provides a set of application guidelines to be used during clinical trials.

Our method answers to the following questions that health care researchers should have before adopting a health care dataset for clinical trials.
\begin{itemize}
    \item Q1. Is there any data bias in the dataset? 
    \item Q2. Which ones are they and how to measure them?
    \item Q3. How to reduce their impact?
\end{itemize}
Several actors will benefit from this method. First, health care researchers, who mainly exploit medical data to carry out clinical trials; they also represent the target users of our method. Secondly, physicians and patients, since they are directly or indirectly involved in the application of results from clinical trials, and in the usage of automated systems for medical diagnosis based on machine learning techniques. Finally, ethics committees will benefit from having guidelines to apply in the health care research context, to extend the ethical principles of medicine also to new, not yet widespread computer-based tools. 

The data management pipeline defined in~\cite{dataManagementPipeline} delineates five main stages of the lifecycle of decision support systems:  requirements, system design, construction, testing, and maintenance. The method we propose will be used after data have been extracted from health care datasets, at the beginning of the construction phase and before the analysis phase.
This temporal position allows our method to be applied after data have been---at least partially---collected, but before the construction of the analysis algorithm. In fact, in our opinion, clinicians need to modify data before the beginning of the construction phase.

The method uses the data bias identified in~\cite{biasHealthCare} and described in Section~\ref{sec:soa}, where a list of data bias in the context of health care are defined. The method we are proposing considers bias in training data, since bias in model design and related to clinicians and patients cannot be analyzed from a computer scientist perspective, due to the lack of measurement strategies. In particular, considered data bias are: \emph{minority, missing data, informativeness} and \emph{selection bias}. 

\subsection{Input/Output}

The input of the method consists of a dataset and information on its application context. The method targets reasonably large datasets that are going to be used in machine learning analytics for evidence-based medicine.
The output of the method consists of a list of biases that have been identified in the data source, their quantification through appropriate metrics, and possible mitigation actions to solve them.
This output allows to: \textit{(i)} inform the researcher about the level of fairness associated to the input dataset by using a variety of bias metrics; \textit{(ii)} increase awareness about the limits of the outcome using the input dataset in clinical trials and consequently, the risks connected to its application; \textit{(iii)} have adequate information for further uses of the datasets, that can be stored in metadata; \textit{(iv)} choose a mitigation strategy that, by applying it, can produce a cleaned version of the input data source.

\subsection{Steps}
\begin{figure}[t]
  \centering
    \includegraphics[width=0.8\textwidth]{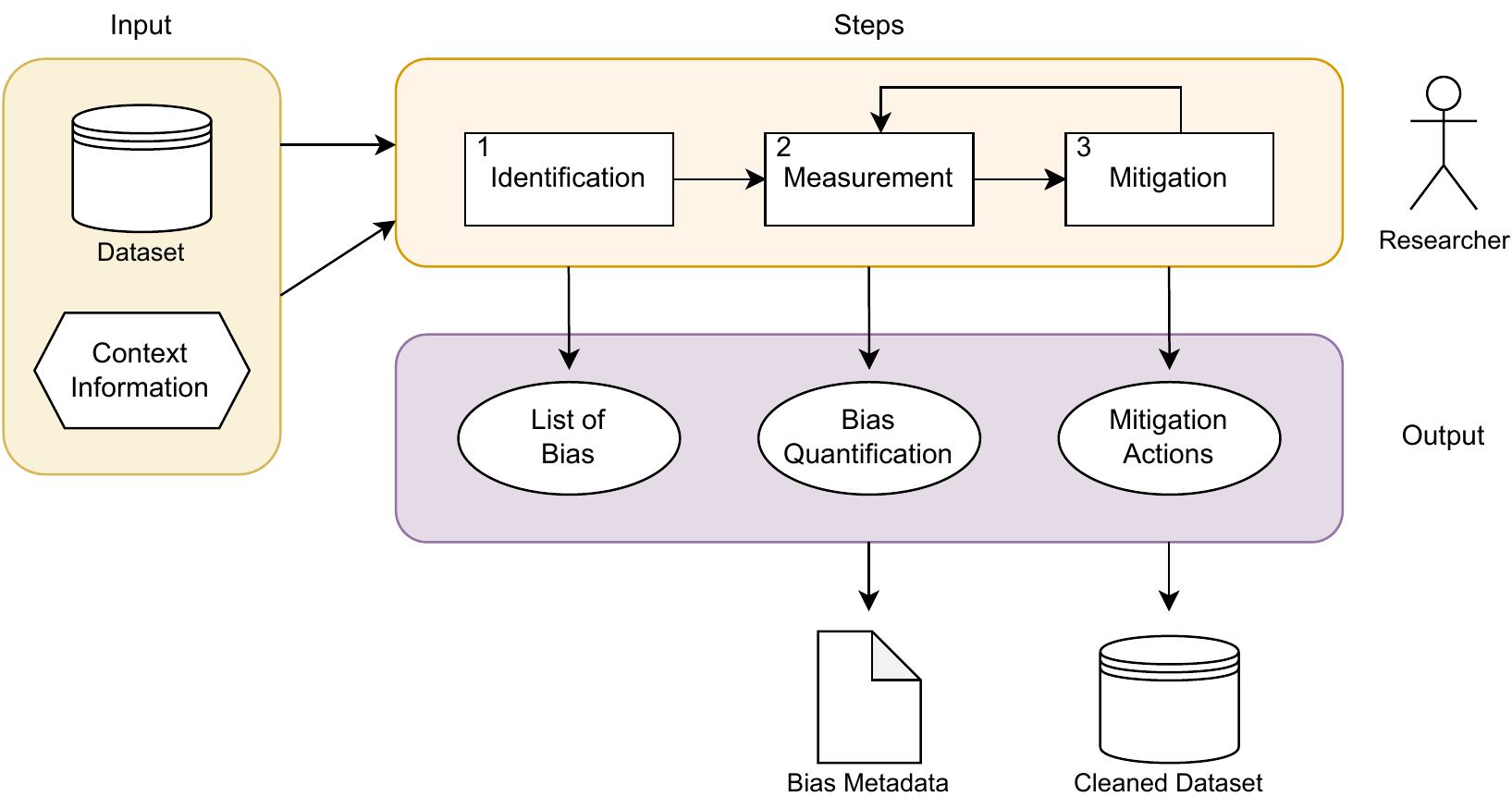}
    \caption{The pipeline of the proposed method.}
    \label{fig:framework}
\end{figure}

Figure~\ref{fig:framework} shows the pipeline of the method proposed in this paper. All steps, described below, will be executed by the health care researcher.

\begin{enumerate}
    \item \emph{Bias Identification} includes the study and analysis on the presence of different types of data bias.
    \item \emph{Bias Measurement}, where each data bias is mapped with its corresponding measurement techniques.
    \item \emph{Bias Mitigation}, where each data bias is associated to possible mitigation strategies.
\end{enumerate}
The rest of this section provides a description of each step, while the next section details each bias, the respective metrics and mitigation actions.

\subsubsection{Bias Identification}
This step consists in identifying data bias that may affect the dataset, and that need to be mitigated if present. Possible data bias are: minority, missing data, informativeness and selection bias. For each of them, we report the definition and a list of questions that the researcher should answer to acquire knowledge of the specific bias, in particular on the risks deriving from their presence.
For this reason, the method we propose takes the form of guidelines for health care researchers to perform bias analysis on health data sources, to understand the risks connected to their usage and better know how to mitigate them. 

\subsubsection{Bias Measurement}
This step consists in quantifying the identified data bias, in order to understand the magnitude of the possible impact on the results of the clinical trial. For each data bias we give an overview of the most suitable metrics, taken from the Data Science Ethics literature, to recognize and quantify it. 
The results of this step can be stored in metadata associated to the dataset, to be reused for future clinical trials. In fact, it is common for a dataset to be used for additional clinical trials~\cite{knoppers2014international,van2015bbmri}. Therefore, it is important to save the previously extracted information on data bias for further analyses not originally planned. To do so, our method stores information on the presence and magnitude of bias. 

\subsubsection{Bias Mitigation} 

The last step of the pipeline consists in choosing a mitigation action based on the previously used metric, the context and the goal of the clinical trial. The choice of the mitigation action is strictly related to the context and the goal of the trial, since each action potentially has different consequences on the dataset. 
The best solution does not always consist in adopting a mitigation action that modifies the dataset: there are cases, such as breast cancer, in which the disease has a different incidence, so the dataset is naturally skewed towards a part of the population.
In fact, sometimes the bias is not relevant for the goal of the clinical trial, so the questions in the first step should guide the user in this step as well; on the contrary, when the bias is undesirable or important, the user should perform a mitigation action.
In this work, we propose only pre-processing actions, allowing health care researchers to modify the dataset before the beginning of the construction phase. The mitigation actions presented are taken from Computer Science and Data Science Ethics literature.
Additionally, the application of a mitigation action needs to be carefully considered, since lowering a value of a metric to mitigate data bias, may raise other metrics values of the same or different data bias. For this reason, the method we propose is iterative: after the mitigation step, the user applies the measurement step again in order to evaluate the cleaned dataset. 
However, we will not further discuss this aspect since it is not the scope of this paper.

\section{Bias Identification and Measurement}\label{sec:bias}

This section explores the bias introduced in the previous Section~\ref{sec:method}, giving the definition, the measurement and the possible mitigation strategies for each one.

\subsection{Minority Bias}
\paragraph{Definition.} The protected groups may have insufﬁcient numbers of patients for a model to learn the correct statistical patterns~\cite[p.3]{biasHealthCare}.

\paragraph{Identification.} During this step, it is important to focus on the attributes and features that are in the dataset. The following questions help health care researchers to identify this data bias.
\begin{itemize}
    \item Which attributes and features are present in the dataset?
    \item Which are more relevant for the study?
    \item Are there protected attributes for the study? Are gender or ethnicity present in the dataset?
\end{itemize}

A group can be identified by specifying one or more characteristics of the attributes present in the dataset. Some examples of groups are: females, males, Caucasian, Black, Asiatic, Caucasian females, Black females, etc. If at least one of these characteristics is protected, the entire group is also protected. 

After answering the questions, the health care researcher starts combining the information given by the context and the goal of the clinical trial, with the identified groups. If she concludes that, given the clinical trial, the dataset is not affected by minority bias, because, for example, there is only one group, then the risks and the issues illustrated in Section~\ref{sec:context} are prevented. For this reason, the identification of different groups, particularly the protected ones, is fundamental. 

\paragraph{Measurement.} To measure the number of people in the (protected) group and, consequently, the minority bias, we identify three different measurements: density, diversity and fairness. Each one studies different aspects of data representation and distribution.

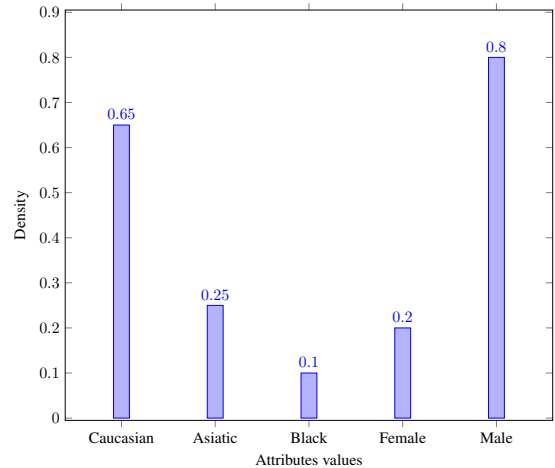
\begin{figure}[t]
\centering 
\scalebox{.6}{
\begin{tikzpicture}
\begin{axis}
[  
    ybar,  
    enlargelimits=0.15,  
    ylabel={Density}, 
    xlabel={Attributes values},  
    symbolic x coords={Caucasian, Asiatic,Black,Female,Male}, 
    xtick=data,  
     nodes near coords, 
    nodes near coords align={vertical},  
    ]  
\addplot coordinates {(Caucasian,+0.65) (Black,0.1) (Asiatic,0.25) (Male,0.8) (Female,0.2) };  
\end{axis}  
\end{tikzpicture} }
\caption{Density plot}
\label{tab:densityPlot}
\end{figure}

First, \emph{Density} is the degree to which different entities occur in the dataset. 
It is the number of occurrences of a value~\cite{naumann2004completeness}, so it represents the easiest metric to study. 
Figure~\ref{tab:densityPlot} shows the density for a simple example where there are two attributes: ``sex" has two values ``Female" and ``Male" and ``ethnicity" has three values ``Caucasian", ``Black" and ``Asiatic". It is evident from the plot that the minorities are females, Black people and Asiatic people.

\begin{figure}[t]
\centering
\begin{minipage}{.5\textwidth}
  \centering
  \includegraphics[width=0.75\linewidth]{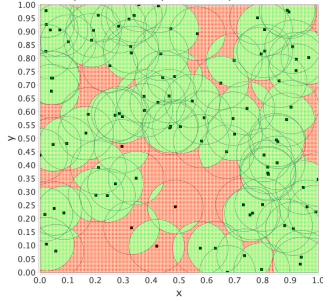}
  \vspace{-1.5mm}
  \caption{Diversity Plot}
  \label{fig:diversityPlot}
\end{minipage}%
\begin{minipage}{.5\textwidth}
  \centering
  \vspace{2mm}
  \includegraphics[width=.85\linewidth]{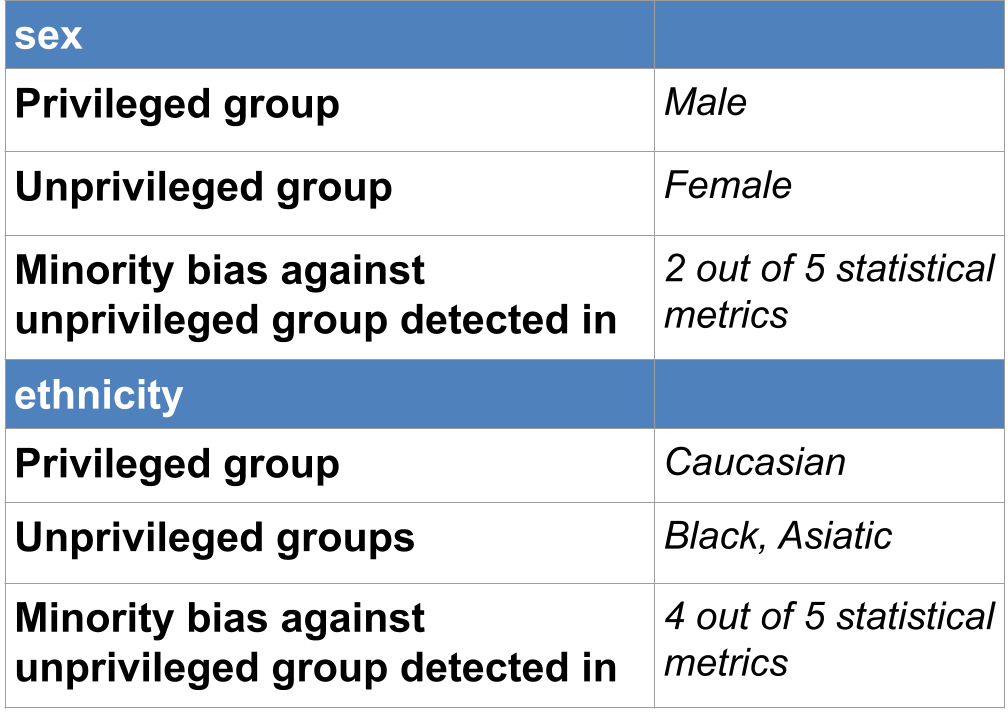}
  \vspace{2mm}
  \caption{Fairness Plot}
  \label{fig:fairnessPlot}
\end{minipage}
\end{figure}

Another measurement is \emph{diversity} that ``ensures that different kinds of objects are represented in the data"~\cite[p.1]{Drosou2017DiversityIB}. The most popular metrics for diversity are based on the notions of distance or coverage~\cite{Drosou2017DiversityIB}. 
A system that defines and studies coverage is given in \cite{asudeh2021identifying}, briefly described in Section \ref{sec:soa}. The authors developed efficient algorithms to identify uncovered regions of data. Figure~\ref{fig:diversityPlot}, taken from~\cite{asudeh2021identifying}, shows a sample dataset with two attributes $x_1$ and $x_2$. Every tuple in the dataset is represented by a black dot, and the uncovered region is colored in red. A green zone is a region that has a sufficient number of samples, while a red zone evidences a lack of elements and, consequently, the presence of minority bias.

To measure \emph{fairness}, the most used metrics are statistical ones~\cite{verma2018fairness}: they evaluate the predictions, given by a classification model based on the dataset, compared with the actual instances of the dataset.
An example of fairness analysis is given in Figure~\ref{fig:fairnessPlot}. In this example we selected five statistical metrics and we analyzed the outcome of the predictions for the different groups. It is worth noticing that the metrics detect minority bias against unprivileged groups, namely ``Female", ``Black" and ``Asiatic" groups. There are tools off-the-shelf that can provide this analysis, such as~\cite{bellamy2018ai}, briefly described in Section \ref{sec:soa}. 

\paragraph{Bias Mitigation.} Different techniques can be adopted to mitigate minority bias.
The choice depends on the goal of the clinical trial and on the metric chosen by the health care researcher. Indeed, if the clinical trial only focuses on the majority groups, without the goal of studying the behavior of minorities, there is no need to change the group distributions, and the researcher should not modify the dataset. Otherwise, she should perform a mitigation action. For space limitation, we present the ones described in Section~\ref{sec:soa}.

If the researcher is interested in density and diversity, the mitigation action to apply is~\cite{asudeh2019assessing}, that determines the least amount of data that the researcher must add to solve the lack of representation previously discovered. If she is interested in fairness, the dataset can be modified using one of the pre-processing techniques presented in~\cite{bellamy2018ai}, such as \emph{reweighing}, that weights the examples in each (group, label) combination differently to ensure fairness before classification, or \emph{optimized pre-processing}, that learns a probabilistic transformation to modify the features and the labels in the data.

\subsection{Missing Data Bias}
    
\paragraph{Definition} Data may be missing for protected groups in a nonrandom fashion, which makes an accurate prediction hard to render~\cite[p.3]{biasHealthCare}.

\paragraph{Identification}
``A missing value is a value that exists in the real world, but is not available in a data collection"~\cite[p.7]{batini2009methodologies}. The following questions may help health care researchers to identify this data bias.
\begin{itemize}
    \item Are there any missing values? If so, why?
\end{itemize}

It is important to understand the reason why a value is missing. In fact, missing values are very frequent in questionnaires for medical scenarios~\cite{ibrahim2012missing}. 
There are three main types of missing values: \emph{(i)} missing completely at random, i.e., the value does not depend on observed attributes or missing value, \emph{(ii)} missing at random, i.e., it depends on observed attributes, \emph{(iii)} missing not at random, when the value that is missing depends on the value itself \cite{little2019statistical}.

To identify this data bias, the user has to concentrate on the last type of missing value. It is critical to understand the types of missing values that the dataset may have, because the user has to understand the reason why the value is missing. Unfortunately, rarely researchers can determine the type of missing value solely based on the data at hand, so context information must be considered as well~\cite{ibrahim2012missing}.

\paragraph{Measurement} In large datasets, identifying missing values can be difficult and often requires domain knowledge. Completeness, defined as ``the ratio between the number of not null values over the total number of values"~\cite[p.19]{batini2009methodologies}, can be used to measure the amount of missing values. It can be computed on values, on tuple, or on attributes~\cite{batini2016data}. Given the example used for the minority bias, Table~\ref{tab:MissingValue} shows a completeness analysis with the percentage of missing values for each attribute value.

\begin{table}[b]
\caption{Completeness analysis}
\label{tab:MissingValue}
\centering
\begin{tabular}{p{25mm} r}
  \toprule
  \emph{Attribute value} & \emph{\% of missing values} \\ \midrule
    Black & 15\% \\
    Caucasian & 5\% \\
    Asiatic & 30\% \\
    Female & 25\% \\
    Male & 0\% \\
  \bottomrule
\end{tabular}
\end{table}

\paragraph{Bias Mitigation.} Three main methods can be used to mitigate the presence of missing values~\cite{ibrahim2005missing}:
\emph{(i)} \emph{Deletion}, in which the researcher discards all the data with a missing values;
\emph{(ii)} \emph{Single Imputation} in which the researcher converts the missing values into a new value, for example substituting the mean, the mode or other statistical values;
\emph{(iii)} \emph{Model-Based}, in which the researcher assign a new value, based on a model created on the rest of the dataset. The researcher should decide the best strategy that yields the least biased estimations in the results, taking into account the context and the goal of the clinical trial.
Moreover, fixing all missing values may require a considerable amount of effort. Therefore, tuples can be marked based on utility contribution, and prioritized based on the research goals. Additionally, to improve cost/effectiveness, the researcher may consider fixing only attributes that can contribute more to the prediction~\cite{batini2009methodologies}.

\subsection{Informativeness Bias}

\paragraph{Definition} Features may be less informative to render a prediction in protected groups~\cite[p.3]{biasHealthCare}.

\paragraph{Identification} The following questions help health care researchers to identify this data bias.
\begin{itemize}
    \item Which features are more important for predicting the result?
    \item Are there features for which more information is needed?
\end{itemize}
To understand which features are used to predict the outcome and how much they are important, the researcher has to analyze the prediction algorithm.
Given an input dataset, the output of an algorithm mainly depends on the technique adopted.

\paragraph{Measurement}
In the machine learning field, advanced techniques such as neural networks create very complex models, and the process that lead to results is not easy to understand; for this reason, they are said to be opaque. The purpose of explainability is to make explicit the interactions among the model, the learning technique adopted to produce it, and the data on which it operates; this is relevant both when we want to understand how the model works, and when we want to explain the outcome of the model for a single individual or group.

Thus, it is fundamental to study \emph{Explainable Machine Learning}, which ``highlights decision-relevant parts of the used representations of the algorithms and active parts in the algorithmic model"~\cite[p.3]{holzinger2019causability}.
If we consider a prediction algorithm like logistic regression, to explain the output of the model and consequently measure this data bias, feature importance techniques can be used. In these tasks, the importance of each feature is given by the coefficient that the model associates to each feature during the training phase: these coefficients indicate how much each attribute value impacts the final output. Therefore, in order to evaluate a model with respect to to the informativeness bias, it is sufficient to study and compare the feature coefficients, and to control whether the most important attributes also exhibit unfair behaviors.
This way we can establish whether the learned model is informative for all groups or not.

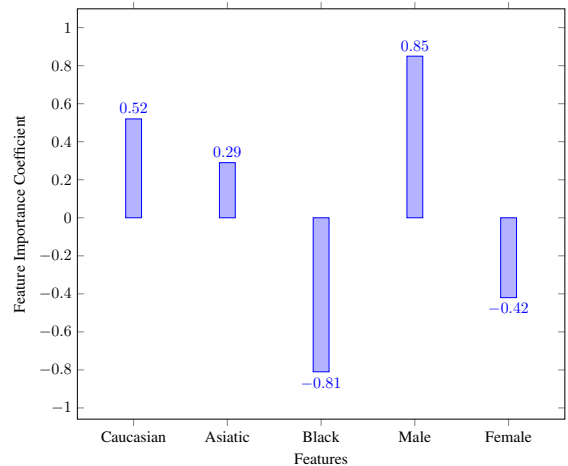
\begin{figure}[t]
\centering 
\scalebox{.6}{
\begin{tikzpicture}
\begin{axis}  
[  
    ybar,  
    enlargelimits=0.15,  
    ylabel={Feature Importance Coefficient}, 
    xlabel={Features},  
    symbolic x coords={Caucasian, Asiatic,Black, Male, Female}, 
    xtick=data,  
     nodes near coords, 
    nodes near coords align={vertical},  
    ]  
\addplot coordinates {(Caucasian,+0.52) (Asiatic,+0.29) (Black,-0.81) (Male,+0.85) (Female, -0.42) };  
\end{axis}  
\end{tikzpicture}}
\caption{Feature importance}
\label{fig:FeatureImportance}
\end{figure}

Figure~\ref{fig:FeatureImportance} shows the feature importance of the previous example: Caucasian, Male and Asiatic group have a higher probability to get a positive outcome with respect to Black and Female that, due to the negative coefficients, will probably receive a negative one.

\paragraph{Bias Mitigation.} Given the goal and the context of the clinical trial, if all the features are informative enough, the researcher can leave the model as it is. However, as said before, not all algorithms are explainable, and feature importance is not always computable.
In this case, two approaches can be adopted:
\emph{(i)} \emph{post-hoc systems} provides local explanations for a specific decision and make it reproducible on demand, instead of explaining the whole system behavior,
\emph{(ii)} \emph{ante-hoc systems} implements algorithms interpretable by design towards specific 
approaches~\cite{holzinger2017glass}, such as linear regression and decision trees.
According to the context information and to the goal of the clinical trial, the researcher can adopt one of these two techniques to solve the problem of informativeness bias. 

\subsection{Selection Bias}

\paragraph{Bias Definition} 
The training data may not be representative of the population, or the deployment data may differ from the training data. 
\cite[p.3]{biasHealthCare}.

\paragraph{Identification} In this step, the researcher answers to the following questions.
\begin{itemize}
    \item Are there training and testing phases for the study?
    \item Has the model been evaluated? If yes, how?
\end{itemize}
Given the whole available data, the procedure to construct a data analysis algorithm is the following: in a first phase, named ``training phase", the model is built using a portion of the data, then in the second phase named ``testing phase" the model is evaluated with the remaining data. 
In some cases, a part of the available data is reserved to develop the model before production: this is called “validation set" or “development set”. For the researcher it is useful to understand how these sets are produced, and whether they are similar and representative of the population.

\begin{figure}[t]
\centering 
\scalebox{.65}{
\begin{tikzpicture}
\begin{axis}[ylabel={Count}, 
    xlabel={Feature Y}, every axis plot post/.append style={
  mark=none,domain=-2:3,samples=50,smooth}, 
  axis x line*=bottom, 
  axis y line*=left, 
  enlargelimits=upper] 
  \addplot[dashed][blue] {gauss(0,0.5)};
  \addplot {gauss(1,0.75)};
\end{axis}
\end{tikzpicture}
}
\caption{Training and test sets distributions}
\label{tab:TrainTestDistribution}
\end{figure}
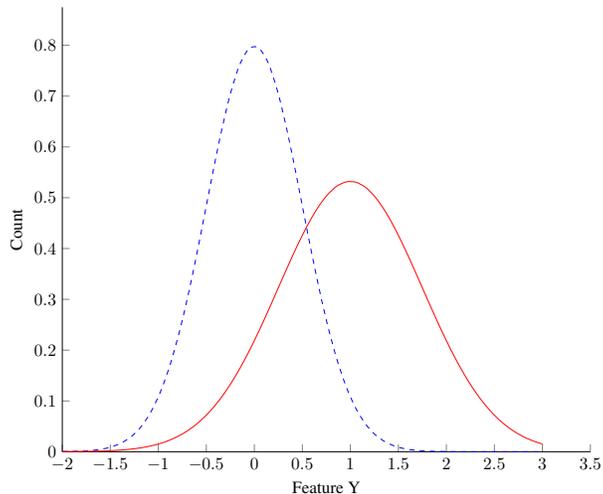

\paragraph{Measurement} 
Given the sets used for training, testing and the eventual validation, in order to measure this type of data bias, the researcher analyzes and compares the distributions of these sets. 
A general \emph{Data Exploration} of the sets based on plots, summary statistics and correlation analysis is highly recommended to understand the shape of the data. Figure~\ref{tab:TrainTestDistribution} plots the training and test sets distributions, with respect to a feature $Y$, from the example described above. The plot suggests that the two distributions are not similar, so in this case the training set and the test set are not equally representative of the population. 

To precisely measure the difference between two data distributions and to compare them, \emph{statistical tests} can be adopted. In general, a statistical test defines a null hypothesis H0 (in our example H0 is ``the two distributions are equal"), an alternative hypothesis H1 (in this case ``the two distributions are different"), and a significance level $\alpha$, i.e. the probability of rejecting H0. The sampling distribution of the statistical test is computable either exactly or approximately, and the value obtained is the p-value. If the p-value is less than or equal to the chosen significance level $\alpha$, then H0 is rejected in favor of H1; otherwise, if the p-value is greater than $\alpha$, the null hypothesis is accepted. Usually, values of $\alpha$ such as 0.05 or 0.01 are chosen. 

\paragraph{Bias Mitigation.}
If the measurement step showed some inconsistencies, a mitigation procedure is needed.
The split between training and test set should be performed avoiding selection bias.
The most common option is to assign two third of the data for training and one third for testing. However, for small or “unbalanced” datasets, samples might not be representative. A more advanced option is to use \emph{K-fold cross-validation}~\cite{zaki2020data}: after splitting data into $K$ subsets of equal size, each subset is used for testing one at a time, with the remaining portion of data being used for training. 
Standard method for evaluation is to use ``repeated K-fold cross-validation", in which the K-fold cross-validation is repeated multiple times. Usually $K$ is chosen to be 5 or 10. The special case, when $K = n$, is called “leave-one-out cross-validation", where the testing set comprises a single point and the remaining data is used for training purposes~\cite{zaki2020data}.

\begin{table}[hb]
    \centering
    \caption{Data bias summary table}
    \label{tab:bias-summary}
    \begin{tabular}{l @{\hskip 8mm} l @{\hskip 8mm} l}
        \toprule
        \textit{Type of data bias} & \textit{Measurement} & \textit{Mitigation} \\
        \midrule
        \multirow{4}{*}{Minority} & Density & \multirow{2}{*}{Diversity enhancement} \\ 
        \vspace{1mm}
        & Diversity \\
        & \multirow{2}{*}{Fairness} & Reweighing \\ & & Optimized pre-processing \\
        \midrule
        \multirow{3}{*}{Missing value} & \multirow{3}{*}{Completeness} & Deletion \\
        & & Single imputation \\
        & & Model-based methods \\
        \midrule
        \multirow{2}{*}{Informativeness} & Explainable ML & Ante-hoc systems \\
        & Feature importance & Post-hoc systems \\
        \midrule
        \multirow{2}{*}{Selection} & Data exploration & \multirow{2}{*}{K-fold cross-validation} \\  & Statistical tests \\
        \bottomrule
    \end{tabular}
\end{table}

\section{Validation}\label{sec:validation}
A preliminary validation of our method consisted in semi-structured interviews with subjects working in the health care sector. We interviewed 5 subjects: \emph{(i)} an oncologist and experienced health care researcher; \emph{(ii)} a philosopher with expertise in bioethics working in a research institute; \emph{(iii)} an experienced imaging and neuro-anatomy technician working in a research laboratory; \emph{(iv)} a mathematician and philosopher, expert in data analysis and health care ethics; \emph{(v)} a physician and pharmacologist working in a research institute and member of ethics committee.

In these interviews, we described our method, being as impartial as possible, and asked the following questions:
\begin{itemize}
    \item Is the temporal placement of the method correct?
    \item Is the list of data bias complete?
    \item Is data bias correctly defined?
    \item Is the list of metrics complete for each data bias?
    \item Do the metrics correctly represent the data bias?
    \item Is the list of mitigation actions complete?
    \item Is the pipeline functional to the objectives of the method?
    \item Are the outputs of the method valuable for the health care researchers?
\end{itemize}

All subjects answered positively to the questions, confirming the validity of the method and underlining its importance for health care researchers and datasets. Through their answers, the subjects highlighted relevant aspects for our research such as: the seriousness of the consequences of data bias in clinical trials, the importance of contextual information and the goal in clinical trials in the application of the method, the need of a common procedure for identifying, measuring and possibly mitigating data bias.
Furthermore, subjects highlighted that, for the effective usage of the method, a collaboration with a computer scientist becomes fundamental, considering that a health care researcher may not be familiar with mitigation actions. This is especially true since not all of the mitigation actions are available off-the-shelf or easily usable, due to the complexity of understanding them. 

From these valuable interactions we have been able to enrich and improve our research, even correcting and adapting the method to the demands and needs of the researchers.

\section{Conclusion}\label{sec:conclusion}

Clinical trials based on big data and machine learning techniques, such as data lakes and data mining, allow to analyze larger and larger datasets obtaining precise results through efficient procedures, leading to new and more reliable therapies.
However, it has been shown that the complexity of their operations significantly reduces the transparency of the results obtained from their use. Therefore, it is of paramount importance to analyze and measure data bias in health care datasets, to raise the trustworthiness of clinical trials and avoid relying on erroneous results.

This paper proposes a method that guides health care researchers in achieving this objective with a pipeline that defines three steps for the identification of data bias, their measurement and mitigation. The method has been validated with semi-structured interviews that underlined the importance of the method and its outputs.

Future work will include: 
\emph{(i)} further analyze the relation between data bias and mitigation actions, improving the questions provided to health care researchers; 
\emph{(ii)} structure metadata for health care datasets, and integrate them in our method; 
\emph{(iii)} conduct experiments on real health care datasets using our method; 
\emph{(iv)} build a software tool that supports our method.

\section*{Acknowledgments} 
This work has been partially supported by the Health Big Data Project (CCR-2018-23669122), funded by the Italian Ministry of Economy and Finance and coordinated by the Italian Ministry of Health and the network Alleanza Contro il Cancro.

\bibliographystyle{unsrt}  
\bibliography{references}  

\end{document}